\documentclass[preprint,amsmath,aps,10pt,superscriptaddress,letterpaper,tightenlines,showkeys]{revtex4}
\usepackage{bm}
\usepackage{stmaryrd}
\usepackage{amsmath}
\usepackage{amssymb}
\usepackage{graphicx}
\usepackage{textcomp}
\usepackage{calrsfs}
\usepackage{yfonts}
\usepackage{color}
\usepackage{amsthm}
\usepackage{lipsum}
\newcommand{\haf}{{\frac{1}{2}}}
\newcommand{\intf}{{\int_0^{\infty}\,}}
\newcommand{\la}{{\langle}}
\newcommand{\ra}{{\rangle}}
\newcommand{\omj}{{\omega_j}}

\newcommand{\omk}{{\omega_k}}
\newcommand{\p}{{\partial}}
\newcommand{\bmu}{{\boldsymbol{\mu}}}
\newcommand{\bxi}{{\boldsymbol{\xi}}}
\begin{document}
\title{On oscillator-bath system: Exact propagator, Reduced density matrix and Green's function}

\author{A. Refaei}
\email{refaei@iausdj.ac.ir}
\affiliation{Department of Physics, Sanandaj branch, Islamic Azad University, Sanandaj, Iran.}

\author{F. Kheirandish}
\email{fkheirandish@yahoo.com}
\affiliation{Department of Physics,
Faculty of Science, The University of Isfahan,
Isfahan, Iran.}
\begin{abstract}
The exact form of quantum propagator of a quantum oscillator interacting with a bosonic bath consisting of $N$ distinguished quantum oscillators with different frequencies is obtained in the Heisenberg picture. Reduced density matrix for oscillator is obtained. The kernel or Green's function connecting the initial density matrix of the oscillator to the density matrix in an arbitrary time is obtained and its connection to Feynman-Vernon influence functional is discussed. Weak coupling regime and squared mean values for position, momentum and energy of the oscillator are obtained in equilibrium.
\end{abstract}
\keywords{Propagator; Bosonic-bath; Density-matrix; Memory function; Green's function}
\maketitle
\section{Introduction}
The theory of open quantum systems has been extensively studied and developed after the seminal works reported in \cite{Mag,Fey,Ull,Cal}. This theory has found applications in fields like  statistical mechanics, chemical physics \cite{1,2,3,4,5}, condensed matter \cite{10,11,12}, quantum optics \cite{13,14,15,16}, quantum information and atomic physics \cite{17}.
The dynamics of an open quantum system can be described in terms of generalized master equations \cite{18,19} and Langevin equations \cite{20,21} and due to the complexity of correctly incorporating different physical parameters, various approximations such as weak-coupling regime, Markovian limit, high temperature, or the initial factorizing condition are usually invoked. This approximations are not valid for open quantum systems in the low temperature regime or in the presence of initial correlations between the system and its environment or in the case of driven non-equilibrium quantum systems \cite{Pachon}.

The effects of environmental degrees of freedom on the system can be investigated with the method of Feynman-Vernon influence functional \cite{Fey} by integrating over environment variables within the context of the closed-time-path formalism \cite{22,23,24,25}.

In the present work, using symmetry and initial condition properties of quantum propagator, the exact form of quantum propagator of a quantum oscillator interacting with a bosonic bath consisting of $N$ distinguished quantum oscillators with different frequencies is obtained in the Heisenberg picture. Knowing the propagator of the total system, reduced density matrix for oscillator is obtained. The kernel or Green's function connecting the initial density matrix of the oscillator to the density matrix in an arbitrary time is obtained and its connection to Feynman-Vernon influence functional is discussed.

\section{Lagrangian}
The total Lagrangian of the oscillator-bath system can be written as \cite{K1}
\begin{equation}\label{L}
 L=\haf m \dot{x}^2-\haf m \omega^2 x^2+\sum_{j} (\haf \rho\dot{X}_j^2-\haf \rho\,\omj^2 X_j^2)+\sum_j f_j \dot{X}_j \,x,
\end{equation}
where the second term is a collection of independent harmonic oscillators with different frequencies and the last term represents the interaction between the main oscillator (system) and the bath oscillators through a linear coupling with coupling constants $f_j$. The conjugate canonical momenta for the system and bath oscillators are defined by
\begin{eqnarray}\label{pP}
  p &=& \frac{\p L}{\p \dot{x}}=m\dot{x}, \\
  P_j &=& \frac{\p L}{\p \dot{X}_j}=\rho\dot{X}_j+f_j x,
\end{eqnarray}
respectively. The total system is now quantized by imposing equal-time commutation relations
\begin{eqnarray}\label{CR}
  && [x,p] = i\hbar,\\
  && [X_j,P_k] = i\hbar\,\delta_{jk}.
\end{eqnarray}
Hamiltonian is also defined by
\begin{eqnarray}\label{H}
 H &=& p\dot{x}+\sum_j P_j \dot{X}_j-L,\nonumber\\
   &=& \frac{p^2}{2m}+\haf m\omega^2 x^2+\sum_j \bigg[\frac{(P_j-f_j x)^2}{2\rho}+\haf\rho\,\omega^2_j \, X^2_j\bigg].
\end{eqnarray}
From Hamiltonian (\ref{H}) and using Heisenberg equations, One can obtain the equations of motion for the system and bath oscillators as
\begin{equation}\label{eqx}
\ddot{x}+\omega^2 x=\frac{1}{m}\sum_j f_j\,\dot{X}_j,
\end{equation}
and
\begin{equation}\label{eqX}
\ddot{X}_j+\omj^2 X_j=-\frac{f_j}{\rho}\,\dot{x}.
\end{equation}
respectively. Equation (\ref{eqX}) has the formal solution
\begin{equation}\label{Xj}
 X_j (t)=X_j^N (t)-\frac{f_j}{\rho}\int_0 ^t dt'\, G_j (t-t')\, \dot{x}(t'),
\end{equation}
where
\begin{equation}\label{XN}
  X_j^N (t) = \cos(\omj t)\, X_j ^N (0)+\frac{\sin(\omj t)}{\rho\,\omj}\,P_j ^N (0),
\end{equation}
is the homogeneous solution which due to the large number of independent bath oscillators and their unknown initial conditions can be interpreted as a noise field. The Green's function of the equation (\ref{eqX}) is defined by
\begin{equation}\label{G}
  G_j (t-t') = \frac{\sin[\omj (t-t')]}{\omj}\,\theta(t-t'),
\end{equation}
which is a retarded Green's function and $\theta(t-t')$ is the Heaviside step function.
Inserting (\ref{Xj}) into (\ref{eqx}) leads to
\begin{equation}\label{ex}
 \ddot{x}+\omega^2 x(t)+\frac{d}{d t}\int_0^t dt'\,\gamma(t-t')\, \dot{x}(t')=f^N (t),
\end{equation}
where
\begin{equation}
 f^N (t)=\frac{1}{m}\sum_j f_j\, \dot{X}_j ^N (t)=\frac{1}{m}\sum_j f_j\, \bigg[\frac{1}{\rho}\cos(\omj t)\, P_j ^N (0)-\omega_j\,\sin(\omj t)\, X_j ^N (0)\bigg],
 \end{equation}
is the scaled noise force. The memory function or susceptibility of the medium or bosonic-bath is defined by
 \begin{equation}\label{gama}
 \gamma(t-t')=\frac{1}{m\rho}\sum_j f_j ^2 \,G_j (t-t')= \frac{1}{m\rho}\sum_j \frac{f_j^2}{\omj}\,\sin[\omj (t-t')]\,\theta(t-t').
\end{equation}
We can rewrite (\ref{gama}) in terms of the spectral density function $g(\omega)$ as
\begin{equation}\label{gama2}
 \gamma(\tau)=\intf d\omega\,g(\omega)\,\sin(\omega\tau),\,\,\,\,\tau=t-t'>0,
\end{equation}
where
\begin{equation}\label{gomega}
 g(\omega)=\frac{1}{m\rho}\sum_j \frac{f^2_j}{\omega}\,\delta(\omega-\omega_j).
\end{equation}
Equation (\ref{gama2}) is a sine Fourier transformation and can be inverted, leading to
\begin{equation}\label{g2}
 g(\omega)=\frac{2}{\pi}\int_0^\infty d\tau\,\gamma(\tau)\,\sin(\omega\tau).
\end{equation}
which can be rewritten in terms of the imaginary part $(\Im)$ of the Fourier transform of the memory function
\begin{equation}\label{gama3}
  \tilde{\gamma}(\omega)=\int_0^\infty d\tau\,\gamma(\tau)\,e^{i\omega\tau},
\end{equation}
as
\begin{equation}\label{coupling}
  g(\omega)=\frac{2}{\pi}\,\Im[\tilde{\gamma}(\omega)].
\end{equation}
Note that if the bosonic bath is a collection of harmonic oscillators with continuum frequency (as is the case) then we use the substitution
\begin{equation}\label{continuum}
 \sum_j f^2_j \,h(\omega_j)\rightarrow\int_0^\infty d\omega\,f^2(\omega)\,h(\omega),
\end{equation}
where $h(\omega)$ is an arbitrary function of $\omega_j$. In this case from (\ref{coupling}) we will find
\begin{equation}\label{cont}
  f^2(\omega)=\frac{2m\rho\omega}{\pi}\,\Im[\tilde{\gamma}(\omega)].
\end{equation}
Transition from discrete to continuum case can be achieved easily trough out the paper, in this case vectors, matrices and their inverse should be considered as states and operators over the semi-infinite interval frequency $[0,\infty)$.

Taking the Laplace transform of (\ref{ex}) leads to
\begin{equation}\label{14}
  s^2 \tilde{x} (s)-s x(0)-\dot{x}(0)+\omega^2 \tilde{x}(s)+s\tilde{\gamma}(s)(s\tilde{x}(s)-x(0))=\tilde{f}^N (s),
\end{equation}
therefore,
\begin{equation}\label{xs}
  \tilde{x}(s)=\frac{s+s\tilde{\gamma}(s)}{s^2+\omega^2+s^2\tilde{\gamma}(s)}\,x(0)+\frac{1}{s^2+\omega^2+s^2\tilde{\gamma}(s)}\,\dot{x}(0)+
  \frac{\tilde{f}^N (s)}{s^2+\omega^2+s^2\tilde{\gamma}(s)}.
\end{equation}
Let us define for simplicity
\begin{equation}\label{alfa}
  \alpha (t) = L^{-1} \bigg[\frac{s+s\tilde{\gamma}(s)}{s^2+\omega^2+s^2\tilde{\gamma}(s)}\bigg],
\end{equation}
\begin{equation}\label{beta}
  \beta (t) = L^{-1} \bigg[\frac{1}{s^2+\omega^2+s^2\tilde{\gamma}(s)}\bigg],
\end{equation}
\begin{equation}\label{delta}
  \delta_j (t) = L^{-1}\bigg[\frac{s}{(s^2+\omj^2)(s^2+\omega^2+s^2\tilde{\gamma}(s)}\bigg],
\end{equation}
\begin{equation}\label{eta}
  \eta_j (t) = L^{-1}\bigg[\frac{\omj}{(s^2+\omj^2)(s^2+\omega^2+s^2\tilde{\gamma}(s)}\bigg],
\end{equation}
note that $\delta_j (t)=\dot{\eta}_j (t)$. Now taking the inverse Laplace transform of (\ref{xs}) and using (\ref{alfa}$-$\ref{eta}) we find
\begin{equation}\label{x}
  x(t)=\alpha(t)\,x(0)+\beta(t)\,\frac{p(0)}{m}+\frac{1}{m}\sum_j f_j\, [\frac{1}{\rho}\,\delta_j (t) P_j ^N (0)-\omj\eta_j (t) X_j ^N (0)].
\end{equation}
We will make use of this equation in the next section to find the total quantum propagator.
\section{Propagator}
In this section we find the quantum propagator of total system using properties of propagator in the framework of Heisenberg approach. In Heisenberg picture the time evolution of oscillator position operator is given by
\begin{equation}\label{P1}
  x(t)=U^\dag(t) x(0) U(t)\Rightarrow U(t)\,x(t) =x(0)\,U(t),
\end{equation}
the matrix elements of the last equality in the position space of the total system $|x\ra\otimes|X_1,X_2,\cdots,\ra\equiv|x,\mathbf{X}\ra$ is
\begin{equation}\label{22}
  \la x,\mathbf{X}|U(t)\,x(t)|x',\mathbf{X}'\ra=\la x,\mathbf{X}|x(0)\,U(t)|x',\mathbf{X}'\ra,
\end{equation}
now from (\ref{x}) and the definition of propagator $K(x,\mathbf{X},t;x',\mathbf{X}')=\la x,\mathbf{X}|U(t)|x',\mathbf{X}'\ra$ we easily find
\begin{equation}\label{diff}
  \bigg[\beta(t)\frac{\p}{\p x'}+\frac{1}{\rho}\sum_j f_j\,\delta_j (t)\frac{\p}{\p X'_j}\bigg]\,K(x,\mathbf{X},t;x',\mathbf{X}')=-\frac{im}{\hbar}\bigg[x-\alpha(t)x'+\frac{1}{m}\sum_j f_j\,\eta_j  (t)\,\omj\,X'_j\bigg]\,K(x,\mathbf{X},t;x',\mathbf{X}').
\end{equation}
or equivalently
\begin{equation}\label{diffx}
  \bigg[\beta(t)\frac{\p}{\p x'}+\frac{1}{\rho}\sum_j f_j\,\delta_j (t)\frac{\p}{\p X'_j}\bigg]\,\ln K(x,\mathbf{X},t;x',\mathbf{X}')=-\frac{im}{\hbar}\bigg[x-\alpha(t)x'+\frac{1}{m}\sum_j f_j\,\eta_j  (t)\,\omj\,X'_j\bigg].
\end{equation}
This is a partial differential equation that propagator should fulfill in but we need another equation which comes from the time evolution of the bath oscillator positions $X_j (t)$ which we now determine. By taking the Laplace transform of (\ref{Xj}) we find
\begin{equation}\label{Xs}
  \tilde{X}_j (s)=\tilde{X}_j ^N (s)-\frac{f_j}{\rho}\,\tilde{G}_j (s)(s\tilde{x}(s)-x(0)),
\end{equation}
where for $t>0$, we have $\tilde{G}_j (s)=L[\sin(\omj t)/\omj]=1/(s^2+\omj^2)$. By inserting (\ref{xs}) and Laplace transform of (\ref{XN}) into (\ref{Xs}), we find
\begin{eqnarray}\label{26}
 \tilde{X}_j (s) &=& \frac{\omega^2\,f_j}{\rho(s^2+\omj^2)(s^2+\omega^2+s^2\tilde{\gamma}(s))}\,x(0)-
 \frac{s f_j}{\rho m (s^2+\omj^2)(s^2+\omega^2+s^2\tilde{\gamma}(s)))}\,p(0)\nonumber\\
 &+& \frac{1}{\rho(s^2+\omj^2)}\sum_k \bigg[\delta_{jk}-\frac{f_j\,f_k\,s^2}{m\rho(s^2+\omega_k^2)(s^2+\omega^2+s^2\tilde{\gamma}(s))}\bigg]\,P_k ^N (0) \nonumber\\
 &+& \frac{s}{(s^2+\omj^2)}\sum_k \bigg[\delta_{jk}+\frac{f_j\,f_k\,\omega_k^2}{m\rho(s^2+\omega_k^2)(s^2+\omega^2+s^2\tilde{\gamma}(s))}\bigg]\,X_k ^N (0),
\end{eqnarray}
and by taking the inverse Laplace transform, we find
\begin{eqnarray}
  X_j (t) &=& \frac{\omega^2 f_j}{\omj\rho}\,\eta_j (t)\,x(0)-\frac{f_j}{\rho m\omj}\,\dot{\eta}_j\,p(0)\,\nonumber\\
  &+& \sum_k \bigg[\mathcal{M}_{jk} (t)\,X_k ^N (0)-\frac{1}{\rho\omk^2}\,\dot{\mathcal{M}}_{jk} (t) P_k ^N (0)\bigg],
\end{eqnarray}
where for notational simplicity we have defined the matrices
\begin{equation}\label{Mjk}
 \mathcal{M}_{jk}=\cos(\omj t)\,\delta_{jk}+Q_{jk} (t)\,\omk^2.
\end{equation}
and
\begin{equation}\label{Qjk}
  Q_{jk}(t)=Q_{kj}(t)=L^{-1}\bigg[\frac{f_j\,f_k \,s}{\rho\, m(s^2+\omj^2)(s^2+\omk^2)(s^2+\omega^2+s^2\tilde{\gamma}(s))}\bigg].
\end{equation}
Now similar to (\ref{P1}) we have
\begin{equation}\label{28}
  X_j (t)=U^\dag(t) X_j(0) U (t)\Rightarrow U(t)\,X_j (t) = X_j (0)\,U(t),
\end{equation}
using
\begin{equation}\label{29}
  \la x,\mathbf{X}|U(t)\,X_j (t)|x',\mathbf{X}'\ra=\la x,\mathbf{X}|X_j (0)\,U(t)|x',\mathbf{X}'\ra,
\end{equation}
and following the same steps that led to (\ref{diffx}), we will find
 \begin{eqnarray}\label{diffX}
  \bigg[\frac{f_j\,\dot{\eta}_j (t)}{\omj}\,\frac{\p}{\p x'}+m\,\sum_k \frac{1}{\omk^2}\dot{\mathcal{M}}_{jk}\,\frac{\p}{\p X'_k}\bigg]\,\ln K(x,\mathbf{X},t;x',\mathbf{X}')
  =\frac{i m\rho}{\hbar}\bigg(X_j -\frac{\omega^2\,f_j}{\omj\rho}\,\eta_j \,x'-\sum_k \mathcal{M}_{jk}\,X'_k \bigg).
 \end{eqnarray}
 The form of quantum propagator can now be determined from equations (\ref{diffx}) and (\ref{diffX}). The right hand side of these partial differential equations suggest that we can assume the following general quadratic form for the logarithm of propagator
 \begin{equation}\label{lnK}
  \ln K(x,\mathbf{X},t;x',\mathbf{X}')=A+B_0 x'+\mathbf{B}\cdot\mathbf{X}'+\haf C_0 x'^2+\haf x'\,\mathbf{C}\cdot\mathbf{X}'+\haf D_{ij}\,X'_i X'_j,
 \end{equation}
 where the coefficients $A,\,B_0,\mathbf{B},\, C_0,\,\mathbf{C}$ and $D_{ij}$ can depend on time and unprimed variables $x,\,\{X_j\}$. Inserting (\ref{lnK}) into equations (\ref{diffx},\ref{diffX}) and matching the coefficients on both sides of these equations we will find
 \begin{eqnarray}
   B_0 &=& -\frac{im}{\hbar\beta}\,x -\frac{im}{\rho\hbar\beta}\,\sum\limits_{jl}\,\dot{\xi}_j\mathcal{N}^{-1}_{jl}\lambda_l, \\
   B_k &=& \frac{im}{\hbar}\,\sum_{l}\mathcal{N}^{-1}_{kl}\,\lambda_l, \\
   C_0 &=& \frac{im\alpha}{\hbar\beta}+\frac{im}{\rho\hbar\beta}\,\sum\limits_{jl}\,\dot{\xi}_j\mathcal{N}^{-1}_{jl}\mu_l, \\
   C_k &=& -\frac{2im}{\hbar}\,\sum_{l}\mathcal{N}^{-1}_{kl}\,\mu_l, \\
   D_{ij} &=&  -\frac{i\rho}{\hbar}\,\sum_{l}\mathcal{N}^{-1}_{il}\mathcal{L}_{lj},
 \end{eqnarray}
 where for notational simplicity we have defined
 \begin{eqnarray}
   \xi_k (t) &=& \frac{f_k\,\eta_k }{\omk}\\
   \lambda_l &=& \rho\, X_l-\frac{x}{\beta}\,\dot{\xi}_l , \\
   \mu_l (t) &=& \omega^2\xi_l(t)+\frac{\alpha}{\beta}\,\dot{\xi}_l, \\
   \mathcal{N}_{jk} &=& \mathcal{N}_{kj}=\frac{m}{\omk^2}\,\dot{\mathcal{M}}_{jk}-\frac{f_j f_k \,\dot{\eta}_j \dot{\eta}_k}{\rho\beta\omega_j \omega_k},\nonumber\\
   &=& -\frac{m}{\omj}\,\sin(\omj t) \,\delta_{jk}+m\dot{Q}_{jk}-\frac{f_j f_k \dot{\eta}_j \dot{\eta}_k}{\rho\beta\omega_j \omega_k},\\
   \mathcal{L}_{ji} &=& m\mathcal{M}_{ji}-\frac{f_j f_i \dot{\eta}_j \eta_i \omega_i}{\rho\beta\omega_j},\nonumber\\
   &=& m\cos(\omj t)\,\delta_{ji}+m\omega_i^2 Q_{ji}(t)-\frac{f_j f_i \dot{\eta}_j \eta_i \,\omega_i}{\rho\beta\omega_j}.
    \end{eqnarray}
By inserting coefficients (42$-$46) into (\ref{lnK}) and making use of the symmetry property $K(x,\mathbf{X},t;x',\mathbf{X}')=K(x',\mathbf{X}',t;x,\mathbf{X})$, we can write the propagator as
\begin{eqnarray}
  K(x,\mathbf{X},t;x',\mathbf{X}') &=& g(t)\, e^{\frac{im}{2\hbar\beta}\{(x^2+x'^2)a(t)-2b(t)\,xx'\}} \\
   &\cdot& e^{-\frac{i\rho}{2\hbar}\{\mathbf{X}'\cdot(\mathcal{N}^{-1}\mathcal{L})\cdot\mathbf{X}'+\mathbf{X}\cdot(\mathcal{N}^{-1}
   \mathcal{L})\cdot\mathbf{X}-2m\,\mathbf{X}' \cdot\mathcal{N}^{-1}\cdot \mathbf{X}\}} \\
   &\cdot&  e^{-\frac{im}{\hbar\beta}\{(x'\mathbf{X}+x\mathbf{X}')\cdot\mathcal{N}^{-1}\cdot\dot{\boldsymbol{\xi}}\}}\,
   e^{-\frac{im}{\hbar}\{(x'\mathbf{X}'+x\mathbf{X})\cdot\mathcal{N}^{-1}\cdot\boldsymbol{\mu}\}},
\end{eqnarray}
where
 \begin{eqnarray}
  a(t) &=& \alpha(t)+\frac{1}{\rho}\dot{\boldsymbol{\xi}}\cdot\mathcal{N}^{-1}\cdot(\omega^2 \boldsymbol{\xi}+\frac{\alpha}{\beta}\dot{\boldsymbol{\xi}}), \\
  b(t) &=& 1-\frac{1}{\rho\beta}\,\dot{\boldsymbol{\xi}}\cdot\mathcal{N}^{-1}\cdot\dot{\boldsymbol{\xi}},
 \end{eqnarray}
and $g(t)$ is a time dependent function which can be determined from the identity
\begin{equation}\label{g1}
  \int dx''\int\prod_{k=1}^N\, dX''_k \,K(x,\mathbf{X},t;x'',\mathbf{X}'')K^*(x',\mathbf{X}',t;x'',\mathbf{X}'')=\delta(x-x')\prod_k \delta(X_k-X'_k),
\end{equation}
up to a phase factor $e^{i\theta}$ as
\begin{equation}\label{g2}
  g(t)=e^{i\theta}\,\sqrt{\frac{m}{2\pi\hbar\beta}}\,\bigg(\sqrt{\frac{m\rho}{2\pi\hbar}}\bigg)^N\,\frac{1}{\sqrt{|\det\mathcal{N}|}}.
\end{equation}
The phase factor can be obtained from the limiting case $f_k=0,\,\forall k$, or simply $\mathbf{f}=0$. In this case the oscillator is not coupled to the bath-oscillators and the form of $g(t)$ is known in this case. Inserting the limits
\begin{eqnarray}
  \lim_{\mathbf{f}\rightarrow 0}\beta(t) &=& \frac{\sin(\omega t)}{\omega},\\
  \lim_{\mathbf{f}\rightarrow 0} \mathcal{N}_{jk}(t) &=& -\frac{m\sin(\omega_j t)}{\omega_j}\,\delta_{jk},
\end{eqnarray}
into (\ref{g2}) leads to
\begin{equation}\label{gfree}
  g(t) = e^{i\theta}\,\sqrt{\frac{m\omega}{2\pi\hbar\sin(\omega t)}}\,\sqrt{\prod_{k=1}^N\frac{\rho\,\omega_k}{2\pi\hbar\sin(\omega_k t)}},
\end{equation}
on the other hand, when there is no interaction between oscillator and the bath we have
\begin{equation}
    g(t)= \sqrt{\frac{m\omega}{2\pi i\hbar\sin(\omega t)}}\,\sqrt{\prod_{k=1}^N\frac{\rho\,\omega_k}{2\pi i\hbar\sin(\omega_k t)}},
\end{equation}
therefore, $e^{i\theta}=e^{-i\frac{(N+1)\pi}{4}}$ and we find $g(t)$ as
\begin{equation}\label{g}
  g(t)=\sqrt{\frac{m}{2\pi i\hbar\beta}}\,\bigg(\sqrt{\frac{m\rho}{2\pi i\hbar}}\bigg)^N\,\frac{1}{\sqrt{|\det\mathcal{N}|}}.
\end{equation}
Finally, the explicit form of the total propagator is
\begin{eqnarray}\label{Propa}
  K(x,\mathbf{X},t;x',\mathbf{X}') &=& \sqrt{\frac{m}{2\pi i\hbar\beta}}\,\bigg(\sqrt{\frac{m\rho}{2\pi i\hbar}}\bigg)^N\,\frac{1}{\sqrt{|\det\mathcal{N}|}}\, e^{\frac{im}{2\hbar\beta}\{(x^2+x'^2)a(t)-2b(t)\,xx'\}} \nonumber\\
   &\cdot& e^{-\frac{i\rho}{2\hbar}\{\mathbf{X}'\cdot(\mathcal{N}^{-1}\mathcal{L})\cdot\mathbf{X}'+\mathbf{X}\cdot(\mathcal{N}^{-1}
   \mathcal{L})\cdot\mathbf{X}-2m\,\mathbf{X}' \cdot\mathcal{N}^{-1}\cdot \mathbf{X}\}}\nonumber \\
   &\cdot&  e^{-\frac{im}{\hbar\beta}\{(x'\mathbf{X}+x\mathbf{X}')\cdot\mathcal{N}^{-1}\cdot\dot{\boldsymbol{\xi}}\}}\,
   e^{-\frac{im}{\hbar}\{(x'\mathbf{X}'+x\mathbf{X})\cdot\mathcal{N}^{-1}\cdot\boldsymbol{\mu}\}}.
  \end{eqnarray}
\section{Weak coupling Regime}
The propagator (\ref{Propa}) is exact and more suitable for numerical calculations but in the limit of weak coupling regime we may obtain an approximate expression up to the first order in coupling functions $\mathbf{f}\equiv (f_1,f_2,\cdots,f_N)$. Ignoring from the second order contributions of coupling functions, we find
\begin{eqnarray}
  \beta(t) &\rightarrow&  \frac{\sin(\omega t)}{\omega},\,\,\, Q_{jk} (t) \rightarrow 0,\,\,\,  a(t) \rightarrow \cos(\omega t),\,\,\, b(t)\rightarrow 1, \\
  \boldsymbol{\lambda} &=& \rho\mathbf{X}-\frac{x}{\beta}\dot{\boldsymbol{\xi}},\,\,\,\boldsymbol{\mu} = \omega^2\,\boldsymbol{\xi}+\frac{\alpha}{\beta}\dot{\boldsymbol{\xi}},\,\,\,
  \eta(t)\rightarrow \frac{\omj\sin(\omega t)-\omega\sin(\omj t)}{\omega(\omj^2-\omega^2)} \\
  \mathcal{M}_{jk}(t) &\rightarrow& \cos(\omj t)\delta_{jk},\,\,\,\mathcal{N}^{-1}_{jk}(t) \rightarrow -\frac{\omj}{m\sin(\omj t)}\,\delta_{jk}, \\
  \mathcal{L}_{jk}(t) &\rightarrow& m\cos(\omj t)\delta_{jk},\,\,\,(\mathcal{N}^{-1}\mathcal{L})_{jk}(t) \rightarrow -\omj\cot(\omj t)\,\delta_{jk}.
\end{eqnarray}
Therefore, up to the first order of coupling functions, the total propagator can be written as
\begin{equation}\label{K1}
 K(x,\mathbf{X},t;x',\mathbf{X}') = K_0 (x,\mathbf{X},t;x',\mathbf{X}')\,e^{\frac{i}{\hbar\beta}\,\sum\limits_{k} \frac{\omk}{\sin(\omk t)}
 \{(xX'_k+x'X_k)\dot{\xi}_k+(xX_k+x'X'_k)(\beta\omega^2\xi_k+\alpha\dot{\xi}_k)\}},
\end{equation}
where $K_0 $ is the propagator in the absence of interaction between oscillator and its environment and has the expected form
\begin{eqnarray}\label{K0}
 K_0 (x,\mathbf{X},t;x',\mathbf{X}') &=& \sqrt{\frac{m\omega}{2\pi i\hbar\sin(\omega t)}}\,e^{\frac{im\omega}{2\hbar\sin(\omega t)}\,[(x^2+x'^2)\cos(\omega t)-2xx']}\nonumber\\
 &\cdot& \prod_{k=1}^N \sqrt{\frac{\rho\omega_k}{2\pi i\hbar\sin(\omega_k t)}}\,e^{\frac{i\rho\omega_k}{2\hbar\sin(\omega_k t)}\,[(X_k^2+X_k'^2)\cos(\omega_k t)-2X_k X_k']}.
\end{eqnarray}
\section{Density matrix}
Having the propagator (\ref{Propa}) we can find the total density matrix in any time. From quantum Liouville equation we have
\begin{equation}\label{den}
  \frac{\p \rho}{\p t}=-\frac{i}{\hbar}\,[H,\rho],
\end{equation}
and since the total Hamiltonian is time-independent we can solve (\ref{den}) as
\begin{equation}\label{den2}
  \rho(t)=e^{-\frac{i t}{\hbar}H}\rho(0)e^{\frac{i t}{\hbar}H}=U(t)\rho(0)U^{\dag} (t).
\end{equation}
Therefore,
\begin{eqnarray}\label{rho}
 \rho(x,\mathbf{X},x',\mathbf{X}',t) &=& \la x,\mathbf{X}|\rho(t)|x',\mathbf{X}'\ra,\nonumber \\
   &=& \la x,\mathbf{X}|U(t)\rho(0)U^{\dag} (t)|x',\mathbf{X}'\ra,\nonumber \\
   &=& \int dx_1 dx_2 d\mathbf{X}_1 d\mathbf{X}_2 \,K(x,\mathbf{X},t;x_1,\mathbf{X}_1)\rho(x_1,\mathbf{X}_1,x_2,\mathbf{X}_2,0)K^{*}(x',\mathbf{X}',t;x_2,\mathbf{X}_2).
\end{eqnarray}
For a given initial state, which is usually a product state $\rho_s\otimes\rho_B$ where $\rho_s$ is an arbitrary density matrix for the oscillator and $\rho_B$ can be chosen for example as a thermal state for the bath, we can find from (\ref{rho}) the total density matrix in an arbitrary time. If we are interested in the time evolution of the reduced density matrix, which is usually the case, then we can tracing out the bath degrees of freedom and find
\begin{eqnarray}\label{reduced}
  \rho_s (x,x',t) &=& tr_B (\rho)=\int d\mathbf{X}\,\rho(x,\mathbf{X},x',\mathbf{X},t),\nonumber\\
  &=&  \int dx_1 dx_2 d\mathbf{X}_1 d\mathbf{X}_2 d\mathbf{X} \,K(x,\mathbf{X},t;x_1,\mathbf{X}_1)\rho_s(x_1,x_2,0) \rho_B (\mathbf{X}_1,\mathbf{X}_2,0)K^{*}(x',\mathbf{X},t;x_2,\mathbf{X}_2),\nonumber\\
  &=& \int \int dx_1 dx_2\, G_{\emph{red}}(x,x';x_1,x_2,t)\,\rho_s (x_1,x_2,0),
\end{eqnarray}
where the kernel or the reduced Green function $G_{\emph{red}}$ is defined by
\begin{equation}\label{redG}
  G_{\emph{red}}(x,x';x_1,x_2,t)=\int d\mathbf{X}_1 d\mathbf{X}_2 d\mathbf{X}\,K(x,\mathbf{X},t;x_1,\mathbf{X}_1)\rho_B(\mathbf{X}_1,\mathbf{X}_2,0)
  K^{*}(x',\mathbf{X},t;x_2,\mathbf{X}_2).
\end{equation}
This kernel is in fact the same factor introduced by Feynman-Vernon in \cite{Fey} known as influence functional obtained from path integral approach. If we define the operator $\mathcal{K}(x,x',t)$ on the Hilbert space of the environment oscillators for real parameters $x,\,x',\,t$ as
\begin{equation}\label{K}
 K(x,\mathbf{X},t;x',\mathbf{X}')=\la \mathbf{X}|\mathcal{K}(x,x',t)|\mathbf{X}'\ra,
\end{equation}
then we can rewrite (\ref{redG}) as
\begin{eqnarray}\label{Gred}
  G_{\emph{red}}(x,x';x_1,x_2,t) &=& \int d\mathbf{X}_1 d\mathbf{X}_2 d\mathbf{X}\,\la \mathbf{X}|\mathcal{K}(x,x_1,t)|\mathbf{X}_1\ra\la \mathbf{X}_1|\rho_B (0)|\mathbf{X}_2\ra\la \mathbf{X}_2|\mathcal{K}^{\dag}(x',x_2,t)|\mathbf{X}\ra,\nonumber\\
  &=& tr[\rho_B (0)\mathcal{K}^{\dag}(x',x_2,t)\mathcal{K}(x,x_1,t)]=\la\mathcal{K}^{\dag}(x',x_2,t)\mathcal{K}(x,x_1,t) \ra_{\emph{eq}}.
\end{eqnarray}
One can show that this recent equation can be rewritten as
\begin{eqnarray}
   G_{\emph{red}}(x,x';x_1,x_2,t) &=& \frac{m}{2\pi\hbar\beta}\,\big(\frac{m\rho}{2\pi\hbar}\big)^N \,e^{-\frac{im}{\hbar}\,\mathbf{q}\cdot\mathcal{N}^{-1}\cdot (x_2 \bmu-\frac{x'}{\beta}\,\dot{\bxi})}\nonumber\\
   &\cdot& e^{\frac{im}{2\hbar\beta}[(x_2^2-x_1^2)\,a+2b\,(x' x_2-x x_1)]}\nonumber \\
   &\cdot& F_B (x'-x,x_2-x_1,t),
\end{eqnarray}
where $F_{B}$ is defined by
\begin{equation}\label{FB}
 F_{B} (x'-x,x_2-x_1,t)=\int d\mathbf{X}\,\rho_B (\mathbf{X},\mathbf{X}+\mathbf{q},0)\,e^{\frac{i}{\hbar}\,\mathbf{X}\cdot[\rho \mathcal{N}^{-1}\mathcal{L}\cdot\mathbf{q}-m\mathcal{N}^{-1}\cdot\mathbf{q}']},
\end{equation}
and
\begin{eqnarray}\label{qp}
  \mathbf{q} &=& \frac{(x_2-x_1)}{\rho\beta}\,\dot{\bxi}-\frac{(x'-x)}{\rho}\bmu, \\
  \mathbf{q}' &=& (x_2-x_1) \bmu-\frac{(x'-x)}{\beta}\,\dot{\bxi}.
\end{eqnarray}
\theoremstyle{definition}
\newtheorem{exmp}{Example}
\begin{exmp}
Let the initial state of the bath be a thermal state given by
\begin{equation}
 \rho_B (\mathbf{X},\mathbf{X}',0)=\prod_{k} \sqrt{\frac{\rho\omega_k}{\pi\hbar}\,\coth(\frac{\omega_k \tau}{2})}\,e^{-\frac{\rho}{2\hbar}
 \sum_k \big[(X_k^2+X_k^{'2})\omega_k\coth(\omega_k \tau)-\frac{2\omega_k}{\sinh(\omega_k \tau)}X_k\,X'_k\big]},
\end{equation}
then from the definition (\ref{FB}) we will find
\begin{equation}\label{W2}
 F_{B} (x'-x,x_2-x_1,t)=2^{\frac{N}{2}}\,e^{\frac{\rho}{4\hbar}\,
 \sum\limits_k \omega_k \coth(\frac{\omega_k \tau}{2})\big[q_k-\frac{i p_k}{2\rho\omega_k \coth(\frac{\omega_k \tau}{2})}\big]^2},
\end{equation}
where $p_k$ is defined by
\begin{equation}\label{pk}
  p_k=\sum\limits_j [\rho\,(\mathcal{N}^{-1}\mathcal{L})_{kj} q_j-m\mathcal{N}^{-1}_{kj} q'_j.
\end{equation}
\end{exmp}
\section{Thermal equilibrium}
In thermal equilibrium the density matrix of the total system is given by
\begin{equation}\label{d1}
  \rho(x,\mathbf{X},x',\mathbf{X}';T)=\frac{1}{Z(T)}\la x,\mathbf{X}|e^{-\frac{1}{k T}\,\hat{H}}|x',\mathbf{X}'\ra=\frac{1}{Z(T)}\,K(x,\mathbf{X},-i\tau ;x',\mathbf{X}')
\end{equation}
where $\tau=i t=\frac{\hbar}{kT}$ and $k$ is the Boltzman constant and the total partition function is given by
\begin{equation}\label{d2}
  Z(T)=\int dx\int d^{N} \mathbf{X}\, \rho(x,\mathbf{X},x,\mathbf{X};T).
\end{equation}
The partition function can be calculated using the formula \cite{zinn}
\begin{equation}\label{d3}
\int d^n x\,e^{-\haf\,\mathbf{x}\cdot A\cdot \mathbf{x}+\mathbf{s}\cdot\mathbf{x}}=(2\pi)^{\frac{N}{2}}\,(\det A)^{-\haf}\,
e^{\haf\,\mathbf{s}\cdot A^{-1}\cdot\mathbf{s}},
\end{equation}
leading to
\begin{equation}\label{Z}
  Z(T) = \frac{1}{i^N\,2^{\frac{N+1}{2}}}\,\frac{m^{\frac{N}{2}}}{\sqrt{\det(\mathcal{L}-m\mathbb{I})}}\,\frac{1}{\sqrt{a-b+\frac{m\zeta}{\rho\beta}}},
\end{equation}
where
\begin{equation}\label{zeta}
  \zeta(t)=(\dot{\bxi}+\beta\bmu)\cdot\mathcal{N}^{-1}(\mathcal{L}-m\mathbf{I})^{-1}\cdot(\dot{\bxi}+\beta\bmu).
\end{equation}
Note that through out the section the time variable $t$ should be replaced with $-i\tau,\,\,(\tau=\frac{\hbar}{k_B\,T})$ in the steady state or long time limit. Now the reduced density matrix of the oscillator can be obtained by tracing over bath variables $\mathbf{X}$ leading to
   \begin{equation}\label{redden}
    \rho(x,x',T)=\sqrt{\frac{m\big[a-b+\frac{m\zeta}{\rho\beta}\big]}{\pi i\hbar\beta}}\,e^{\frac{im}{2\hbar\beta}\,\big\{(x^2+x'^2)(a+\frac{m\zeta}{2\rho\beta})-2(b-\frac{m\zeta}{2\rho\beta})xx'\big\}}.
   \end{equation}
By making use of (\ref{redden}) the mean squared position and momentum of the oscillator can be obtained in a general medium as
\begin{eqnarray}
  \la x^2\ra &=& tr(\rho x^2)=\frac{i\hbar\beta}{2m\big[a-b+\frac{m\zeta}{\rho\beta}\big]}\bigg|_{t=-i\tau}, \\
  \la p^2\ra &=& tr(\rho p^2)=\frac{m\hbar(a+b)}{2i\beta}\bigg|_{t=-i\tau},
\end{eqnarray}
and for energy we find
\begin{equation}\label{E}
  \la H\ra=\frac{\hbar(a+b)}{4i\beta}+\frac{\hbar\omega^2 i\beta}{4\big[a-b+\frac{m\zeta}{\rho\beta}\big]}\bigg|_{t=-i\tau}.
\end{equation}
\section{Conclusions}
In the present work, using the symmetry and initial condition properties of quantum propagator, the exact form of the total propagator of a quantum oscillator interacting with a bosonic bath is obtained in the Heisenberg picture. Knowing the propagator of the total system, reduced density matrix for oscillator is obtained. The kernel or Green's function, connecting the initial density matrix of the oscillator to the density matrix in an arbitrary time is defined and its connection to Feynman-Vernon influence functional is discussed. Weak coupling regime and squared mean values for position, momentum and energy of the oscillator are obtained in equilibrium.

\end{document}